\documentclass{article}
\usepackage[preprint]{neurips_2018}
\usepackage[utf8]{inputenc} 
\usepackage[T1]{fontenc}    
\usepackage{hyperref}       
\usepackage{url}            
\usepackage{booktabs}       
\usepackage{amsfonts}       
\usepackage{nicefrac}       
\usepackage{microtype}      
\usepackage{graphicx} 
\usepackage{subcaption}
\usepackage{natbib}

\usepackage{algorithm}
\usepackage{algorithmic}

\usepackage{hyperref}

\title{Contextualisation of eCommerce Users}
\author{%
  Hassan Elhabbak\\
  Digital Advanced Analytics\\
  Adidas International,Amsterdam\\
  Netherlands, 1101 BA \\
  \texttt{hhappak@gmail.com} \\
  \And
  Beno\^it Descamps  \\
    Digital Advanced Analytics\\
  Adidas International,Amsterdam\\
  Netherlands, 1101 BA \\
  \texttt{benoitdescamps@hotmail.com} \\
  \AND
   Elisabeth Fischer\\
    Data Science Chair\\
  Julius-Maximilians Universit\"{a}t W\''{u}rzburg\\
  Germany, 91074  \\
  \texttt{elisabeth.fischer@informatik.uni-wuerzburg.de} \\
  \And
   Sakis Athanasiadis\\
   Digital Advanced Analytics\\
  Adidas International,Amsterdam\\
  Netherlands, 1101 BA \\
  \texttt{sakisp.athanasiadis@gmail.com} 
}

\begin{document} 
\maketitle
\begin{abstract} 
A scaleable modelling framework for the consumer intent within the setting of e-Commerce is presented. The methodology applies contextualisation through embeddings borrowed from Natural Language Processing. 

By considering the user session journeys throughough the pages of a website as documents, we capture contextual relationships between pages, as well as the topics of the of user visits. Finally, we empirically study the consistency and the stability of the presented framework.

\end{abstract} 

\section{Introduction}
\label{introduction}

Millions of users interact daily on global e-Commerce websites yielding billions of data points. 
Understanding the users’ intent is a key step towards personalising their consumer journeys. 
However, extracting the true user intent from the noise is a daunting task. 
Every single user interaction is unfortunately not necessarily representative of a user's visit. 
Click interactions are known to be noisy and highly biased as shown in the case of recommendation \cite{positionbias}. 
Despite the above, these interactions have been found to be valuable sources of implicit feedback \cite{implicit_feedback_clicks}.
In this work, we take a few steps to address this challenge, aiming to provide a definition of consumer intent within an e-Commerce setting.
Our definition is presented in section (\ref{subsection-framework}). 
We elaborate on our methodology in section (\ref{Methodology}) and 
finally present empirical evidence of consistency and stability for our constructs in section (\ref{section-cluster-intent}). 

\subsection{What are e-Commerce users looking for?}
e-Commerce organizations are faced with the challenge of presenting a vast and often seasonal inventory to a broad and varied audience.
 This inventory ranges from hundreds to tens of thousands of articles and consumers, hopefully, visit the website with a purpose of buying a number of items. 
 Therefore, interactions with the provided content should intuitively indicate their intent.

Various modelling strategies have already been applied as to hedge user 
interactions, ranging from latent modelling to neural networks based approaches \cite{intent_association},\cite{intent_rnn}.
\begin{figure}[t]
\begin{center}
\centerline{\includegraphics[width=100mm]{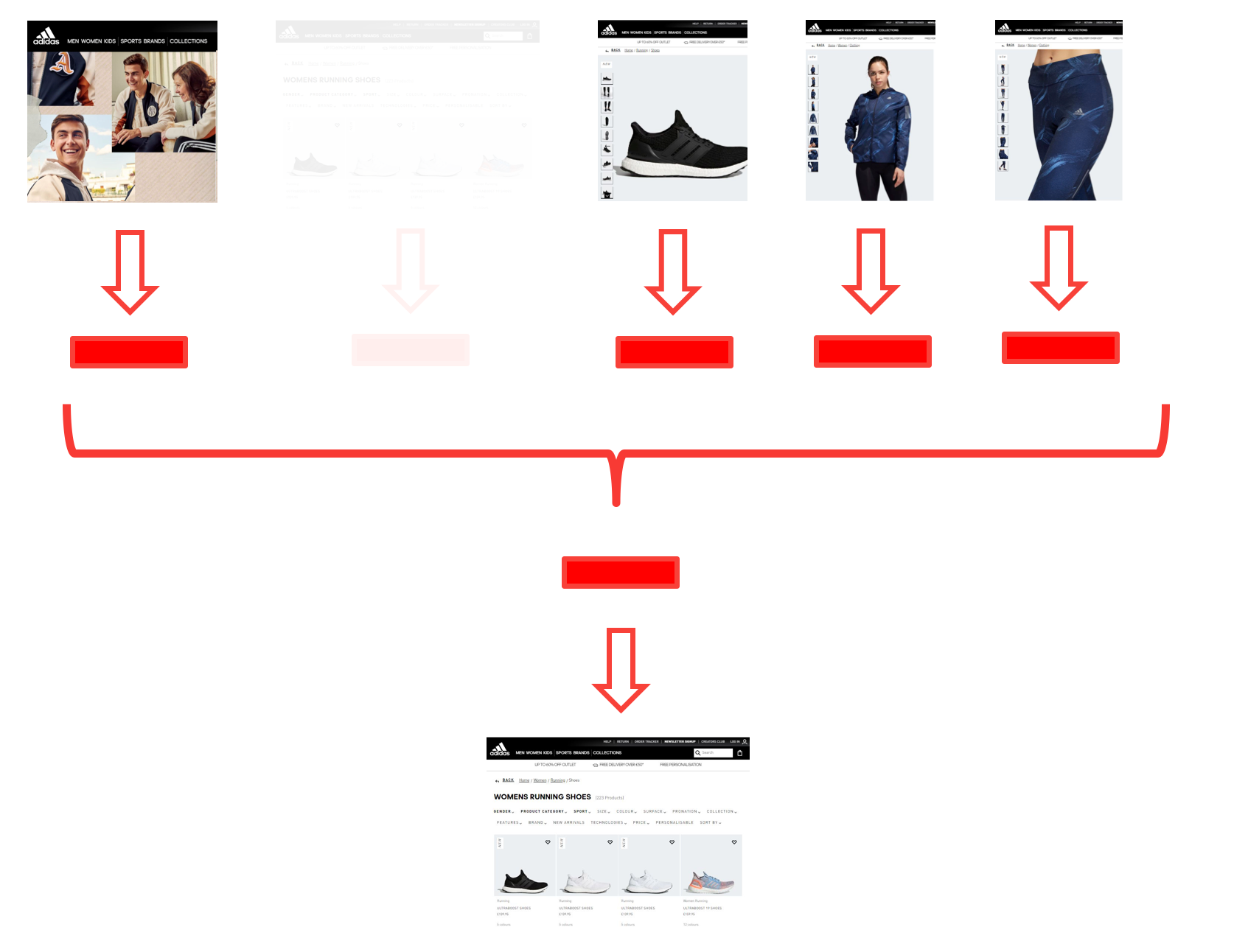}}
\caption{Continuous Bag of Words (CBOW) reframed within a session journey. 
Users navigate through the website in search for a particular set of items. 
The sequence of pages reflect a particular context. 
CBOW determines context by masking a page from this sequence and attempting to predict the latter from the other pages. }
\label{fig-session-cbow}
\end{center}
\end{figure}
Our goal for this work is not to present the most accurate model, but identify 
the smallest key feature space for consumer intent. 

Click interactions are noisy and highly dimensional. Large e-Commerce platforms are built from hundreds of thousands of pages per market. 
 Each identified topic serves a label for segmented targetting strategies.  Designing page content is costly. 
 While fully personalised automated content is still a long way ahead, a segmented approach is quite feasible. 
 Each topic of visit is used as a label for a particular theme of user journey, for which tailored content can be developed.

\section{The user's context}
\subsection{Context in language}

Word representations dates back from 1986 in the seminal work by Rumelhart, Hinton, and Williams 
\cite{Rumelhart1986LearningRB}. Since then optimised modelling strategies have been 
introduced such Continuous Bag of Words (CBOW)\cite{w2v}, Skip-gram \cite{skipgram},  and Glove \cite{glove}.
These word representations,i.e. embeddings, capture similarities between words from the context 
in which they have been observed together.
For an example, the word 'ball' often appears in sport related documents, so does 'football' and 'soccer'. 
Hence,  'ball', 'football' and 'soccer' will be represented (closely) by similar `contextual` embeddings. 

Formally contextual representations are extracted by attempting to optimize the mutual distance between words, 
while correlating with the probability of appearing within the same context. 
For example, in the case of CBOW, the context is modelled by a neural network that is optimised on a target word for a given word context.

Once a contextual representation is defined through words, the next step is understanding a meaningful representation 
for sentences. Encoding sentences from word embeddings has been studied extensively using transformers 
\cite{attention} and deep averaging networks \cite{dan_classification}. 
The latter work showed some good results by simply averaging the word embeddings.

Carrying on with our example, the words ’ball’, ’football’ and ’soccer’ rarely occur in the same context  
along with ’banana’ since word representations tend to form clusters. 
Some clusters can be close to each other as words may represent different topics. 
Mathematically, we couple a distribution of topics with each word.
 In the next sections, we estimate this distribution with Gaussian mixtures.

\subsection{Context in eCommerce}

It should be clear that the previous discussion does not necessarily need to be restricted to language processing. 
Representation learning has been extended and has been under heavy study recently in the context of graphs \cite{node2vec} , \cite{node2vec_zhou}. 
Graph embeddings is a skip-gram method applied to a dataset generated by sampling techniques emphasizing on a particular graph property. 
For our study, we let the users determine the sample of our graph, i.e. the website where each node is a page. 
The consumer sessions when can perceived as paths between the nodes of the graph, generating samples of their intent with respect to the website's offer collection.

Thus, a user’s session is an expression of the user’s intent, similar to how a sentence can be an expression of a need or a feeling.
By treating every interaction from the user along the journey, we construct a sentence which has an inherent grammar and meaning. 
The expected relationship of the semantics of the content,  is captured by maximising the equation above, 
similar to how skip-gram can map the relationship of words and their likely context.

\begin{figure}[t]
\begin{center}
\centerline{\includegraphics[width=150mm]{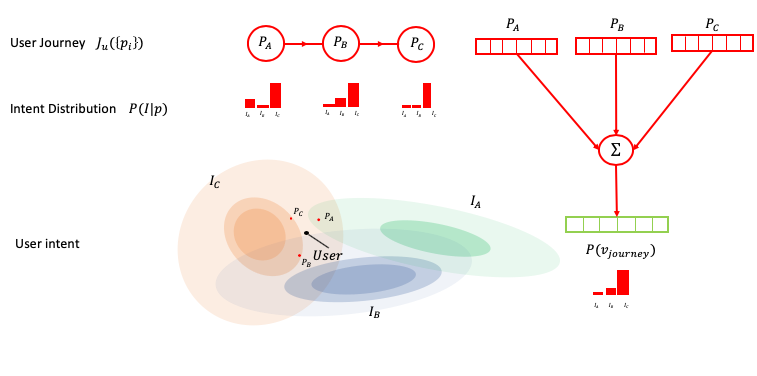}}
\caption{Illustration of the intent modelling framework. The user journey constitutes of a set of ordered pages $\mathcal{J}_u(\{p_i\})$. 
Each page is a distribution of topic of intent. The user intent can be seen as the center of the main cluster determined by the
contextual embeddings of the pages. On inference, the user journey vector $v_{\mbox{\tiny user}}$ is taken to be 
the average of the vector representations of the pages observed within the sessions.
The distribution of the  intent of the user P($v_{\mbox{\tiny user}}$) is computed from the intent distribution of the pages. }
\label{fig-framework}
\end{center}
\end{figure}
Analogous to NLP, (or more general node2vec \cite{node2vec} ), a website is a vocabulary of pages. 
Each page is coupled to a distribution of intents. Users visit a certain number of pages of the website in a particular order.

Formally, let us denote the set of the pages of a website as $\{p_i\}$, and  the set of user intents $\{I_j\}$.  For each page, 
we define an intent distribution $P(I_j|p_i)$. The user journey is represented by an ordered set of pages 
$\mathcal{J}_u(\{p_i\})$. Finally, our goal is to model the probability of the user intent given his/her journey $P(I_j|\mathcal{J}_u(\{p_i\})$.

The estimation of $P(I|p_j) $ consists out of three steps
\begin{enumerate}
 \item Contextualisation of the pages through a vector-representation $v_p$ estimated with methods such as skip-gram
 \item  The distribution of the vector representations $P(v_p) $ is estimated with a mixture of distributions, such as gaussian mixtures \cite{gaussian}. 
   \begin{equation}
   \label{eq-page-distribution}
   P(v_p)  = \sum \limits_{I \in \mathcal{I}} P(I_j|p_i)
     \end{equation}
    Each distribution $P(I_j|p_i)$ is interpreted as the distribution of the page with respect to intent. 
 The intent distribution is thus the vector,
 \begin{equation}
   \label{eq-intent-distribution}
 P(I|p_j)   =\left[P_0(v_p),\dots,P_{\left|\mathcal{I}\right|-1}(v_p) \right]
  \end{equation}
 
\end{enumerate}

\section{Methodology}\label{Methodology}
In this section, we present a methodology which estimates and infers the user intent through their browsing interactions.
We discuss the data preparation, followed by the modelling strategies of the intent 
distributions of the pages (\ref{eq-page-distribution}) and finally how the consumer intent is derived ((\ref{eq-intent-distribution}).
We refer to (\ref{section-cluster-intent}) for a detailed discussion of the assumptions made in this section.
\begin{figure}[t]
\vskip 0.2in
\begin{center}
\centerline{\includegraphics[width=75mm]{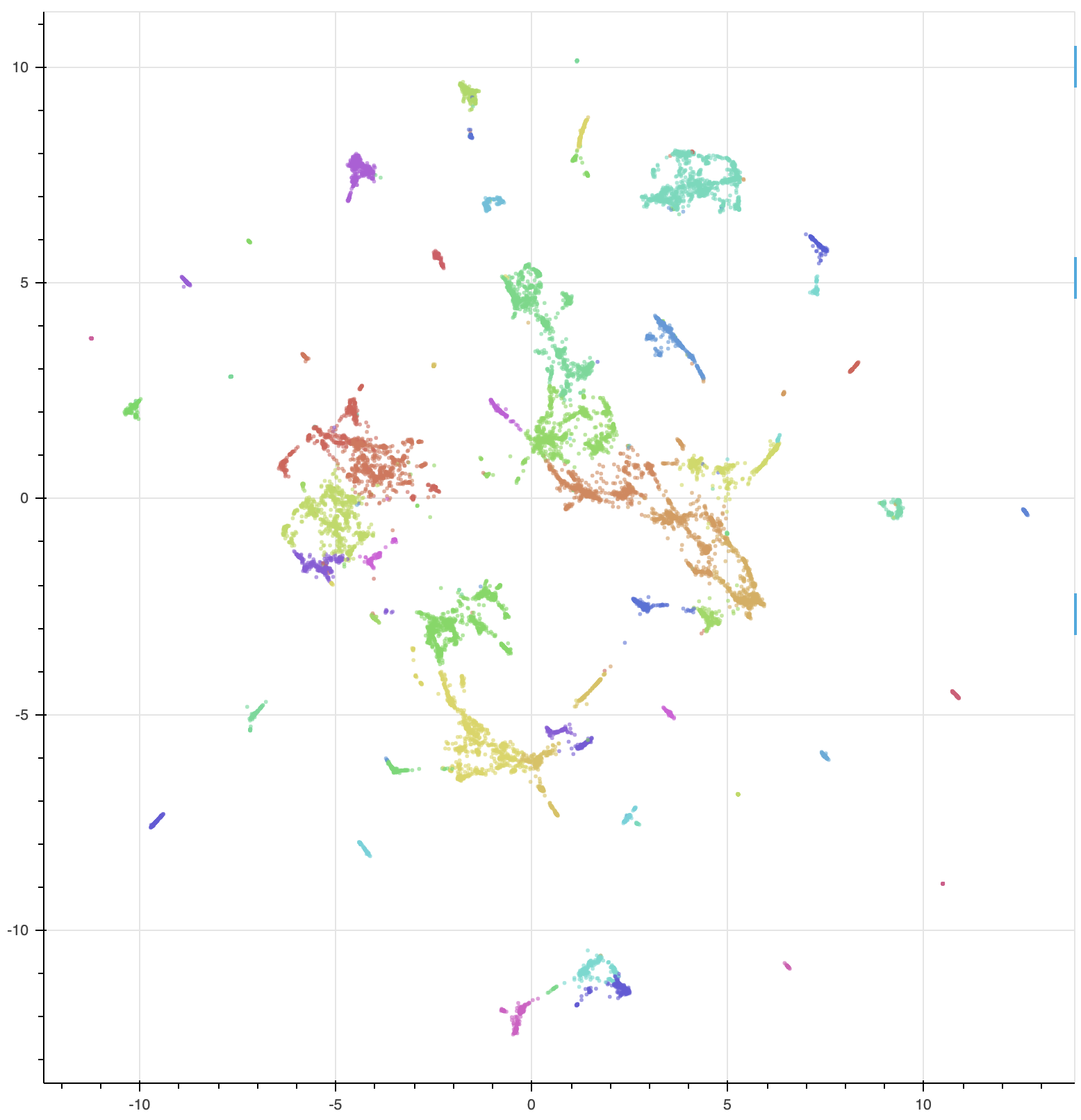}}
\caption{The plot shows the percentage of sessions that included multiple pages with different labeled topics. Majority of sessions contained only one topic.}
\label{fig-umap}
\end{center}
\vskip -0.2in
\end{figure}

\subsection{Framework}\label{subsection-framework}

\subsection{Data}
We process over a year of sessions for a major market. For each session, we collect an ordered list of viewed pages. 
The data consists out of tokens, i.e., webpages interacted by the users . 
The number of tokens considered after processing and simple type grouping in certain cases was close to sixty thousand. 
Short sessions are dropped out of the dataset.
\subsection{Estimation of Page-representation}
The data consisting of a few million observations are fed to a skip-gram algorithm to learn the context of the vocabulary.
 The vocabulary from here on is defined as the (eventually post-processed) the pageID that the observations are constructed with. 
 The output in this case is a dictionary of the embedded vocabulary into vectors.

\subsection{Estimation of User Intent}
\label{subsection-user-intent-methodology}
We make a clear distinction between the estimation and inference of the user-intent.
 In the estimation phase, a model is applied to cluster the vector-representation space of the page-vocabulary. 
 For the inference part, we assume that user sessions live in the same space, reduced to the average vector of the interacted pages.

Gaussian mixtures are chosen to estimate the intent distribution of the page vectors $v$,
  $$P(I|v_p)  =\left[P(I_0|v),\dots,P(I_{\left|\mathcal{I}\right|-1}|v) \right]$$
where each $P_i(v)$ denotes the  probability of the intent $I_J$. 

Let us denote user $u$, observing a sequence of pages $[p_0,\dots,p_s]$. We 
define, an aggregate vector from this sequence, such as the average. 
We apply the gaussian mixture estimate on this session vector aggregate, 
$$  [p_0,\dots, p_s] \rightarrow v_{\mbox{sess}} = \mbox{mean}(v_0,\dots, v_s)\rightarrow v_{\mbox{sess}}  $$
This newly estimated distribution $P(v_p)$ yields the distribution of the membership 
of the user session to each intent topic. The applied aggregation of the session vector can be seen as a basic 
document embedding constructed by the word vectors \cite{dan_classification}.

\begin{figure}[t]
\begin{center}
  \includegraphics[width=75mm]{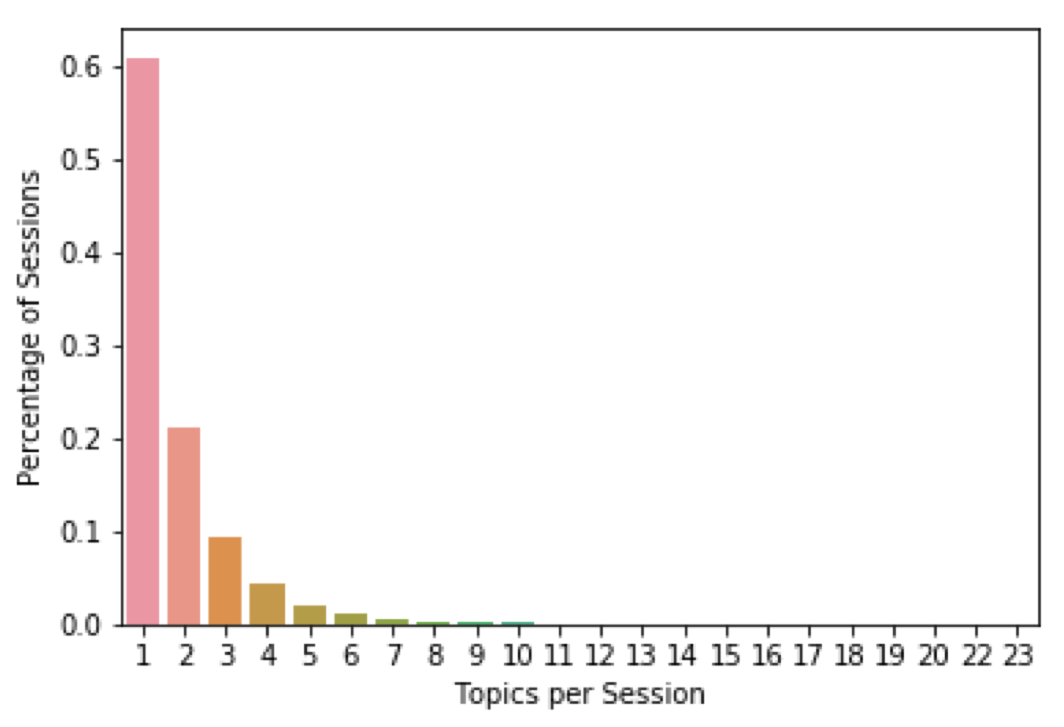}
\caption{The plot shows the percentage of sessions that included multiple pages with different labeled topics. Majority of sessions contained only one topic.}
\label{fig-in-sess-topic-freqs}
\end{center}
\end{figure}
The framework and subsequent modelling strategy presented earlier raises a couple of questions.  
In first instance, we ask whether the observed interactions may not necessarily be a result of users randomly walking through the various marketing funnels of the 
website. If this is the case, users would observe multiple themes of intent during their visit. We need to assert a number of assumptions regarding e-Commerce users. We very these empirically in the next section.

First of all users visit an ecommerce-retailer with a fixed intent. This intent 
  may be an interest in a set of products. As people, combine multiple products 
  and ideally, with proper marketing funnels, explore new items in the inventory, 
  the user's intent within session does not need to be unique. Secondly, a user intent should span over multiple sessions but decays over 
  time as trends and hype in markets change the user's focus. Finally, users retain some consistency in their exploration of the content. This 
   means that not all topics should be connected with each other.
   
As to reflect these properties, following output properties needs to be reflected by our 
models.
\begin{itemize}
  \item If we label the pages with the most likely topic, then the distribution of distinct topics per session should peak around one and decay with the number of topics.
  \item The transition probability between each page's labels should peak for a 
  couple of other labels.
  \item A well-defined session-vector representation should retain some temporal 
  stability over short periods as to reflect the user-interest spanning over 
  multiple sessions.
  \end{itemize}

Skip-gram satisfies the first property as by construction it minimises the mutual distances between contextually similar pages. A clustering algorithm, which retains the metric properties of the vectors captures this property. In the next subsection, we further investigate these empirically.

\subsection{Consistency of Page-representations}
The noisy nature of clicks are a known challenge when modelling click-interactions. It is a fair assumption to note that possibly no pattern could be found from the clicking user. As all pages on a website are typically reachable after a couple of clicks, a website can be practically seen as a dense-graph. If the users behaviour would be similar to a random walk on a dense graph, then no clusters should be observed. 

In figure (\ref{fig-umap}), a U-MAP \cite{umap} is applied on the vector representation of the pages, as to visualise the mutual global distance between each page in two dimensions. We clearly observe multiple clusters of pages.

\begin{figure}[t]
\begin{center}
\centerline{\includegraphics[width=\columnwidth]{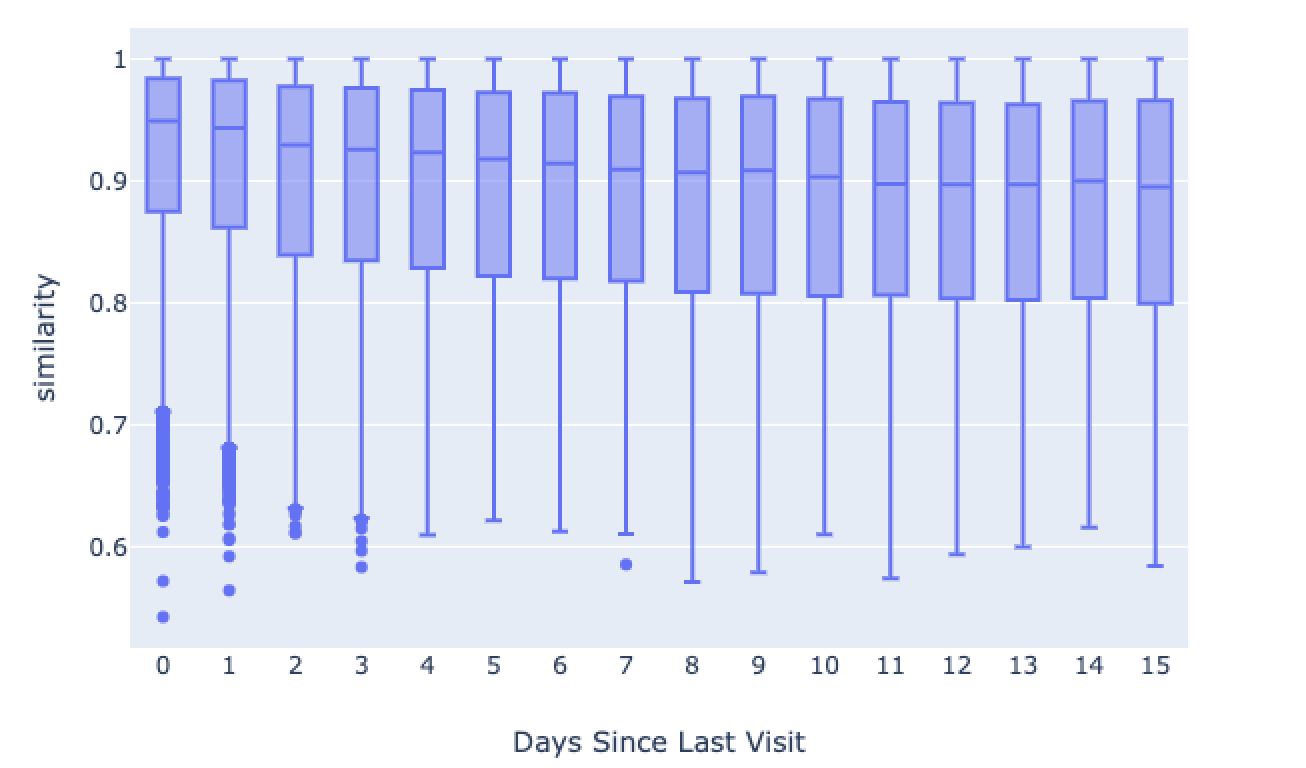} }
\caption{This plot shows the decay of the session similarity of the same returning users given the span of time between the visits. A consistent decay over time is observed.}
\label{fig-session-similarity}
\end{center}
\end{figure}
\section{Clusters in case of intent}
\label{section-cluster-intent}
The inventory of Adidas consists out of three broad categories of items, accessories, shoes and clothing. These categories are reflected within figure  (\ref{fig-umap}) ) as larger clusters englobing multiple clusters. 
\subsection{Stability of the Topics within Session}
The methodology presented earlier \ref{subsection-user-intent-methodology}, does not directly estimate the distribution of the intent from the user 
sessions. One, therefore, needs to ask, whether with this method, user sessions retain some kind of uniqueness in their intent. 
As with sentence encoding, it often turns out that aggregates of the word representation retain consistency. 
This is true as a sentence is constituted of a single topic. We explore this hypothesis empirically, 
by analyzing the frequency distribution of labels within sessions and transition estimates of labels within the same sessions. 

In figure  (\ref{fig-in-sess-topic-freqs}), the distribution of the intent labels within each session is displayed based 
on data from a medium sized market over a period of one year. 
Remarkably, over $60 \%$ of the sessions retain a unique label, while over $80\%$ have upmost 2 labels. 
This supports the initial assumptions of a consistent intent throughough each 
sessions.

\begin{figure}[t]
\vskip 0.2in
\begin{center}
\centerline{\includegraphics[width=\columnwidth]{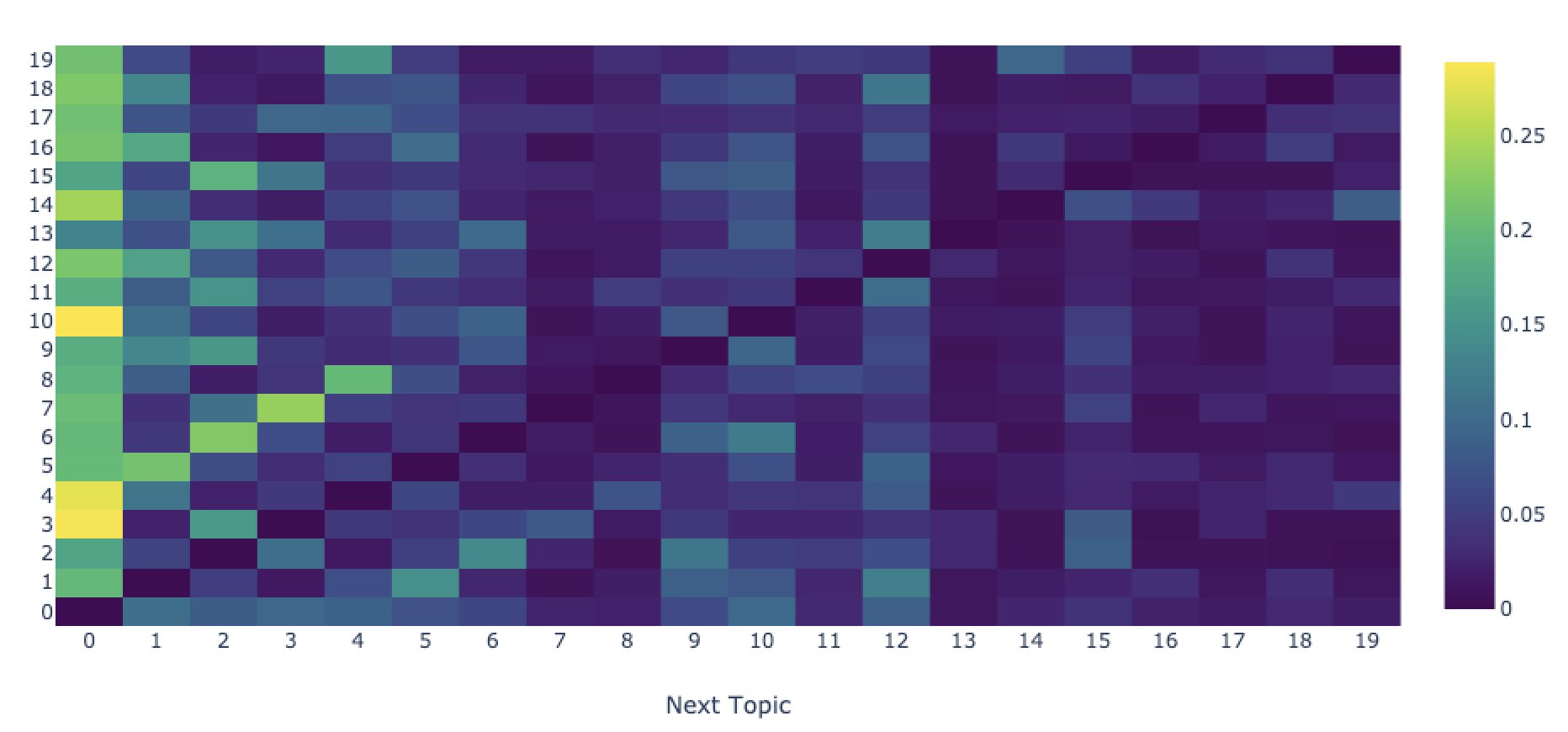}}
\caption{The transition matrix shown here plots probability of transition from one topic to another. Some topics were very popular and end up being combined with other topics more frequently.}
\label{fig-topic-transition}
\end{center}
\vskip -0.2in
\end{figure}

The topic transitions are visualised in figure (\ref{fig-topic-transition}). Most notably, we see that transitions peak at a fixed number of topics. 
This is explained due to the brand-aspect of the service of Adidas. 
There is a high probability that the users at least click on a shoe-related topic during their visit, before finally converging towards their true intent.

\subsection{Stability of the Topics between Sessions}

An analysis by Pinterest showed that purchasing propensity could potential span across multiple sessions, even a couple of weeks apart \cite{pinterest_user_behavior}. 
In the previous, we discussed the stability of the intent within session. For targeting purposes, one cannot always directly act upon the inference of the user-intent.
 It is therefore of interest to analyse the change of intent of user over different sessions.

Due to loss of interest or attention span, the user interest in a particular topic decays and transitions over time. 
In this section, we examine this behaviour.

\begin{figure}

\centering
\begin{subfigure}{0.5\textwidth}
\centering
\includegraphics[width=\linewidth]{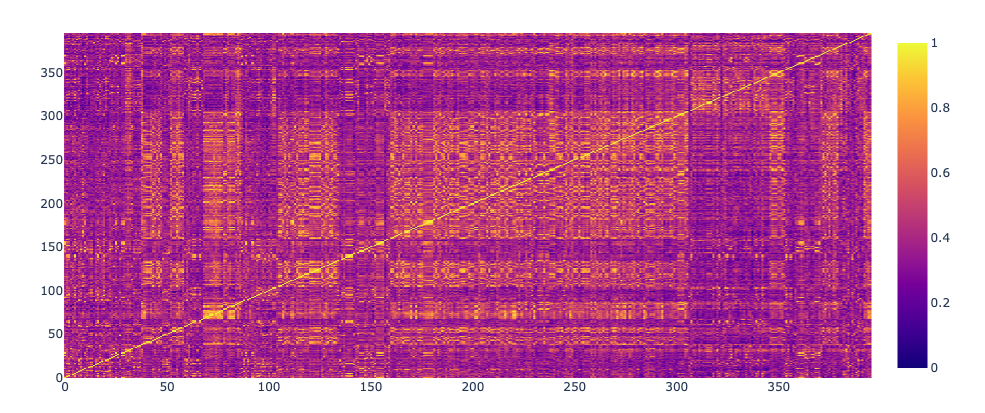}
\caption{Visual Similarity}
\label{label-figure1}
\end{subfigure}%
\begin{subfigure}{0.48\textwidth}
\centering
\includegraphics[width=\linewidth]{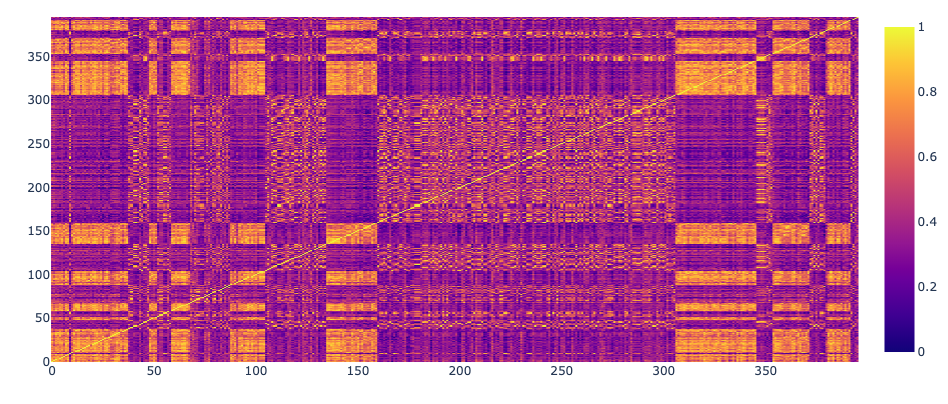}
\caption{Contextual similarity}
\label{label-figure2}
\end{subfigure}
 \caption{Similarity matrices of a sample of product. Notice how blocks of similar article appear both for visual as well as contextual embeddings }
\end{figure}

Given the the vector of a session observed at time $t_0$, described in section 
\ref{subsection-user-intent-methodology}), we compute the similarity $s(v_{\mbox{sess}}(t_0),v_{\mbox{sess}}(t_0)+\Delta t) $
 with the subsequent session-vector $v_{\mbox{sess}}(t_0+\Delta t)$. The result 
 is plot in figure (\ref{fig-session-similarity}),.

\subsection{Visual versus Contextual similarity}
We conclude with a final analysis comparing the page
similarities from a contextual perspective to their similarities from visual content perspective. 
By utilising an Auto-Encoder \cite{autoencoder} model trained on the content images, we are able to estimate visual embeddings as well. 
Auto Encoders are proven to correlate complex image data in order to map out the vector space representing the similarity of the content as viewed by consumers. 
If two pages have similar context, such as articles from the same collection with different colors or slightly different patterns, 
it would be a fair assumption to assume that consumers will end up comparing these pages, hence contextually these pages are close.

As seen in figure (1), the correlation matrix seems to agree on some articles but disagree on others. 
This proved to be very interesting. In the case of our dataset, the products are Adidas articles on Adidas website. 
The contextual model shows that while sometimes soccer shoes can look visually dissimilar, they are still treated contextually the same by users. 
Other times the visual and contextual similarity completely agreed, like in case of clothes that serve the same purpose to users. 

By combining both representations, we can develop a stronger understanding of the user’s intent and purpose, as well as the relationship between the content and the consumers. 

\section{Further Research And Conclusion}

One can argue that the vector representations captured within the modelling 
framework, is restricted to the funnel structure of the ecommerce. Therefore, 
the resulting contextual segments, are a first approximation of intent.

Intent can be seen as a set of interests. While a consumer initially visits the 
website with a particular intent, it does not exclude the existance of additional 
interests.
For future research, we want to look more into the effect of context reframing 
\cite{reframing} on the detected intent for users. Our hypothesis is that when users combine different intents in the same session, the overall intent shifts. 
Understanding how intents combine and interact to form new meanings for the users and the 
business show high potential for simplified targeting solution.

Another interesting direction would be to examine what the consumer intents mean for the product recommendion and product ranking space. 
The information captured could help in tailoring the experience to consumer’s needs and aid in quicker and more accurate training as the feature space becomes more efficient in its representation of information.

In this work, we have reflected on the theory and practice of capturing the intent of a consumer visit on an e-commerce website, and how we can represent it. 
We explored how it could help an e-commerce business tailor their services to consumers and its potential impact to business taxonomy of words. 

\section*{Acknowledgement}
The authors acknowledge financial support from adidas International Marketing B.V. and adidas International B.V. We would like to thank Joshan Meenwa for supporting the initiave. We thank Andras Molnar for sharing the visual embeddings of articles. We are grateful to Alexander Piazza for insightful discussions.

\medskip

\small

\bibliography{context_final}

\begin{thebibliography}{17}
\providecommand{\natexlab}[1]{#1}
\providecommand{\url}[1]{\texttt{#1}}
\expandafter\ifx\csname urlstyle\endcsname\relax
  \providecommand{\doi}[1]{doi: #1}\else
  \providecommand{\doi}{doi: \begingroup \urlstyle{rm}\Url}\fi

\bibitem[Aslanyan \& Porwal(2018)Aslanyan and Porwal]{positionbias}
Aslanyan, G. and Porwal, U.
\newblock Direct estimation of position bias for unbiased learning-to-rank
  without intervention.
\newblock \emph{CoRR}, abs/1812.09338, 2018.

\bibitem[Dempster et~al.(1977)Dempster, Laird, and Rubin]{gaussian}
Dempster, A., Laird, N., and Rubin, D.
\newblock Maximum likelihood from incomplete data via the em algorithm.
\newblock \emph{Journal of the Royal Statistical Society. Series B
  (Methodological)}, 39:\penalty0 1--38, 01 1977.

\bibitem[Grover \& Leskovec(2016)Grover and Leskovec]{node2vec}
Grover, A. and Leskovec, J.
\newblock node2vec: Scalable feature learning for networks.
\newblock \emph{CoRR}, abs/1607.00653, 2016.

\bibitem[Hoque et~al.(2016)Hoque, Ahmed, Lachiche, Leung, and Zhang]{reframing}
Hoque, N., Ahmed, C., Lachiche, N., Leung, C., and Zhang, H.
\newblock Reframing in clustering.
\newblock pp.\  350--354, 11 2016.
\newblock \doi{10.1109/ICTAI.2016.0060}.

\bibitem[Iyyer et~al.(2015)Iyyer, Manjunatha, Boyd-Graber, and
  III]{dan_classification}
Iyyer, M., Manjunatha, V., Boyd-Graber, J., and III, H.
\newblock Deep unordered composition rivals syntactic methods for text
  classification.
\newblock pp.\  1681--1691, 01 2015.
\newblock \doi{10.3115/v1/P15-1162}.

\bibitem[Joachims et~al.(2007)Joachims, Granka, Pan, Hembrooke, Radlinski, and
  Gay]{implicit_feedback_clicks}
Joachims, T., Granka, L., Pan, B., Hembrooke, H., Radlinski, F., and Gay, G.
\newblock Evaluating the accuracy of implicit feedback from clicks and query
  reformulations in web search.
\newblock \emph{ACM Trans. Inf. Syst.}, 25\penalty0 (2), April 2007.

\bibitem[Kingma \& Welling(2014)Kingma and Welling]{autoencoder}
Kingma, D. and Welling, M.
\newblock Auto-encoding variational bayes.
\newblock 12 2014.

\bibitem[Lo et~al.(2016)Lo, Frankowski, and Leskovec]{pinterest_user_behavior}
Lo, C., Frankowski, D., and Leskovec, J.
\newblock Understanding behaviors that lead to purchasing: A case study of
  pinterest.
\newblock In \emph{Proceedings of the 22nd ACM SIGKDD international conference
  on knowledge discovery and data mining}, pp.\  531--540. ACM, 2016.

\bibitem[McInnes \& Healy(2018)McInnes and Healy]{umap}
McInnes, L. and Healy, J.
\newblock Umap: Uniform manifold approximation and projection for dimension
  reduction.
\newblock 02 2018.

\bibitem[Mikolov et~al.(2013{\natexlab{a}})Mikolov, Chen, Corrado, and
  Dean]{skipgram}
Mikolov, T., Chen, K., Corrado, G.~S., and Dean, J.
\newblock Efficient estimation of word representations in vector space.
\newblock \emph{CoRR}, abs/1301.3781, 2013{\natexlab{a}}.

\bibitem[Mikolov et~al.(2013{\natexlab{b}})Mikolov, Sutskever, Chen, Corrado,
  and Dean]{w2v}
Mikolov, T., Sutskever, I., Chen, K., Corrado, G., and Dean, J.
\newblock Distributed representations of words and phrases and their
  compositionality.
\newblock \emph{Advances in Neural Information Processing Systems}, 26, 10
  2013{\natexlab{b}}.

\bibitem[Pennington et~al.(2014)Pennington, Socher, and Manning]{glove}
Pennington, J., Socher, R., and Manning, C.
\newblock {G}love: Global vectors for word representation.
\newblock In \emph{Proceedings of the 2014 Conference on Empirical Methods in
  Natural Language Processing ({EMNLP})}, pp.\  1532--1543, Doha, Qatar,
  October 2014. Association for Computational Linguistics.
\newblock \doi{10.3115/v1/D14-1162}.
\newblock URL \url{https://www.aclweb.org/anthology/D14-1162}.

\bibitem[Rumelhart et~al.(1986)Rumelhart, Hinton, and
  Williams]{Rumelhart1986LearningRB}
Rumelhart, D.~E., Hinton, G.~E., and Williams, R.~J.
\newblock Learning representations by back-propagating errors.
\newblock \emph{Nature}, 323:\penalty0 533--536, 1986.

\bibitem[Sathish \& Patankar(2019)Sathish and Patankar]{intent_association}
Sathish, S. and Patankar, A.
\newblock \emph{Intent Based Association Modeling for E-commerce}, pp.\
  144--156.
\newblock 06 2019.
\newblock ISBN 978-3-030-23280-1.
\newblock \doi{10.1007/978-3-030-23281-8_12}.

\bibitem[Sheil et~al.(2018)Sheil, Rana, and Reilly]{intent_rnn}
Sheil, H., Rana, O., and Reilly, R.
\newblock Predicting purchasing intent: Automatic feature learning using
  recurrent neural networks.
\newblock \emph{CoRR}, abs/1807.08207, 2018.

\bibitem[Vaswani et~al.(2017)Vaswani, Shazeer, Parmar, Uszkoreit, Jones, Gomez,
  Kaiser, and Polosukhin]{attention}
Vaswani, A., Shazeer, N., Parmar, N., Uszkoreit, J., Jones, L., Gomez, A.~N.,
  Kaiser, L., and Polosukhin, I.
\newblock Attention is all you need.
\newblock \emph{CoRR}, abs/1706.03762, 2017.

\bibitem[Zhou et~al.(2018)Zhou, Niu, and Chen]{node2vec_zhou}
Zhou, D., Niu, S., and Chen, S.
\newblock Efficient graph computation for node2vec.
\newblock 05 2018.

\end{thebibliography}
\bibliographystyle{icml2019}

\end{document}